\documentclass[aip,amsmath,amssymb,reprint,onecolumn]{revtex4-2}

\usepackage{graphicx}
\usepackage{dcolumn}
\usepackage{bm}

\usepackage[utf8]{inputenc}
\usepackage[T1]{fontenc}
\usepackage{mathptmx}
\usepackage{etoolbox}

\usepackage{multirow}
\usepackage[dvipsnames]{xcolor}

\makeatletter
\def\@email#1#2{%
 \endgroup
 \patchcmd{\titleblock@produce}
  {\frontmatter@RRAPformat}
  {\frontmatter@RRAPformat{\produce@RRAP{*#1\href{mailto:#2}{#2}}}\frontmatter@RRAPformat}
  {}{}
}%
\makeatother
\begin{document}

\title{Emergence of chirality by multipole interconversion}

\author{Hiroaki Kusunose}
\affiliation{Department of Physics, Meiji University, 214-8571 Japan}
\affiliation{Quantum Research Center for Chirality, Institute for Molecular Science, 444-8585 Japan}
\author{Jun-ichiro Kishine}
\affiliation{Quantum Research Center for Chirality, Institute for Molecular Science, 444-8585 Japan}
\affiliation{Division of Natural and Environmental Sciences, The Open University of Japan, 261-8586, Japan}
\author{Hiroshi M. Yamamoto}
\affiliation{Quantum Research Center for Chirality, Institute for Molecular Science, 444-8585 Japan}
\affiliation{Research Centre of Integrated Molecular Systems, Institute for Molecular Science, 444-8585 Japan}

\date{\today}

\begin{abstract}

A clear understanding of chirality in spin-active electronic states is discussed in order to address confusions about chiral effects recently discovered in materials science. Electronic toroidal monopole $G_0$ can serve as a measure of chirality in this categorization, which can be clearly related to the chiral density operator in the Dirac equation. We extend the concepts of chirality not only to those of materials but also to those of physical fields, and to material-field composites. Additionally, we illustrate specific examples from physics and chemistry that demonstrate the process of acquiring chirality through the combination of seemingly achiral degrees of freedom, which we term the emergence of chirality. Interference among multiple chiralities exhibiting phenomena specific to handedness is also discussed.

\end{abstract}

\maketitle

\section{Introduction}

The term \textit{chirality} was introduced by Lord Kelvin to elucidate the structural properties of molecular isomers whose mirror images cannot be superimposed onto the originals, akin to distinguishing between right and left hands~\cite{Kelvin1904}.
Pasteur, who succeeded in separating a racemic mixture of tartrate into right and left enantiomers from conglomerate crystals (spontaneous resolution) for the first time, also attempted to separate the enantiomers using a magnetic field, but his efforts were in vain~\cite{Barron2012}.
Pierre Curie~\cite{Curie1894} also attempted to generate enantiomeric excess (asymmetry in the population of right/left-handed molecules) by employing collinear electric and magnetic fields based on symmetry viewpoints, a concept classified as \textit{false} chirality due to its time-reversal (T) odd property according to Barron's definition~\cite{Barron2012,Barron2020}.
These efforts have raised a fundamental question: why do chiral molecules found in natural products almost always exhibit enantiopure structures, while molecules synthesized through chemical reactions tend to be racemic?
The origin of such symmetry breaking in the natural molecular world, known as homochirality, remains incompletely understood and is a topic of significant discussion~\cite{Wagnire2007,Blackmond2019}.
Many researchers posit that autocatalytic chirality amplification in biological activity was necessary once finite enantiomeric excess was created by some means (which could include various physical fields, circularly polarized light, a surface of quartz, or even extraterrestrial sources~\cite{Axon1982}).

Chirality also manifests in relativistic quantum theory, playing a crucial role in particle physics.
The Dirac equation in ($3+1$) space-time dimensions involves four $4\times4$ matrices known as the Dirac gamma matrices, and the chirality operator is defined as $\gamma^5\equiv i\gamma^0\gamma^1\gamma^2\gamma^3$, which becomes a conserved quantity for a massless particle corresponding to its handedness~\cite{Sakurai1967}.
There is also a closely related but distinct quantity, helicity, which is the projection of intrinsic angular momentum along the direction of motion. From a symmetry viewpoint, both chirality and helicity satisfy the property of being T-even pseudoscalars, referred to as \textit{true} chirality by Barron.

Solid-state physics offers another realm for exploring the manifestations of chirality.
While it is unclear when the term began to be used in solid-state physics, many types of electronic states are considered chiral, although the original definition of chirality does not always apply.
For example, chiral magnetic soliton lattice~\cite{Dzyaloshinskii1965,Kishine2005,Kishine2015,Togawa2016,Togawa2023} and chiral skyrmions~\cite{Bogdanov1989,Muhlbauer2009,Nagaosa2013} are both truly chiral.
Circularly polarized phonons with finite orbital angular momentum in propagating along the direction parallel or anti-parallel to the angular momentum are also chiral~\cite{Kishine2020,Ishito2023,Ishito2023a,Tsunetsugu2023,Kato2023,Ueda2023,Oishi2024}.
However, \textit{chiral} is sometimes used merely to describe circular motions in systems such as ``chiral'' superconductors\cite{Kallin2016} and 2D ``chiral'' phonons~\cite{Zhang2015}, neither of which align with the definition of chirality.
In the first part of this article, the authors will attempt to categorize these \textit{chirality} in solid-state physics based on symmetry considerations.

In the field of recent physical chemistry, a novel phenomenon associated with chirality has appeared, termed Chirality-Induced Spin Selectivity (CISS).
This effect enables the alignment of electron spins collinearly (either parallel or anti-parallel depending on the molecular handedness) to their momentum upon traversing a chiral molecule~\cite{Ray1999,Naaman2022,Naaman2015}.
CISS has attracted significant attention for its potential in advancing spintronics and spin-related chemistry and physics, owing to its substantial magnitude and ubiquitous occurrence, even in both conductive~\cite{Inui2020,Shishido2021} and insulating~\cite{Ohe2024} inorganic crystals.
Despite ongoing debates regarding the microscopic mechanism, it is noteworthy that structural chirality is intricately linked to electronic chirality, because CISS implies a conversion from structural chirality onto electronic one in the form of helicity.
Historically, there has been a prevailing notion that structural chirality and electronic chirality are unrelated.
Therefore, it is fundamentally crucial to reassess how chirality is unambiguously defined, encompassing electron motion and physical fields in a unified manner.

In our previous paper~\cite{Kishine2022}, the quantum mechanical definition of chirality was given as an electric toroidal (ET) monopole $G_{0}$, i.e., a T-even pseudoscalar, in the framework of symmetry-adapted complete multipole basis~\cite{Kusunose2020,Kusunose2023,Hayami2024,Kusunose2024}.
It was important that any physical quantities in materials and fields can be decomposed into four types of multipole bases at quantum mechanical level, which are categorized according to fundamental symmetries of time-reversal (T), spatial inversion (P), and rotation (R) operations.
The enumeration of (T,P) combinations comprises electric (E), magnetic (M), electric toroidal (ET), and magnetic toroidal (MT) multipoles, which constitute the symmetry-adapted complete basis set.
Each multipole has symbol to express its (T,P) combination and its rank: E, M, ET, and MT monopoles are $Q_{0}$, $M_{0}$, $G_{0}$, $T_{0}$, for example, and their corresponding dipoles are $\bm{Q}$, $\bm{M}$, $\bm{G}$, and $\bm{T}$ (See also Table~\ref{Table1}).
The complete multipole basis enables us to discuss explicitly various forms of chirality in a unified manner, unambiguously distinct from non-chiral quantities.
This aspect is significantly important to consider \textit{emergence of chirality} via multipole interconversion.

First, we will discuss the widely used chirality involving electron spins and/or physical fields in terms of the symmetry-adapted multipole basis.
In particular, time-reversal symmetry is essential in categorizing spin \textit{chirality}.
Therefore, we will discuss how the T-even pseudoscalar $G_{0}$ can be constructed by spin configuration~\cite{Cheong2022} with or without electron motion.
Secondly, we will relate the electronic chirality in weak relativistic regime from the Dirac equation. 
Although it was proved the presence of $G_{0}$ representing a signature of chirality from viewpoint of completeness of the multipole basis, it is now obtained a clear pathway to the relativistic chirality of electron.
On the basis of symmetry-adapted multipole basis, it also becomes possible to discuss emergent chirality using both electronic states and external physical fields by combining various multipoles reflecting their physical states.
Finally, the authors will discuss that the chirality-induced phenomena, i.e., handedness-specific responses, or chiral effects, can be understood in a unified manner by lifting energetically degenerate two chiral states of a concerned system by an interaction with another chiral object and/or field.

\section{Chirality of spin configuration}

\begin{figure}
\includegraphics[width=16cm]{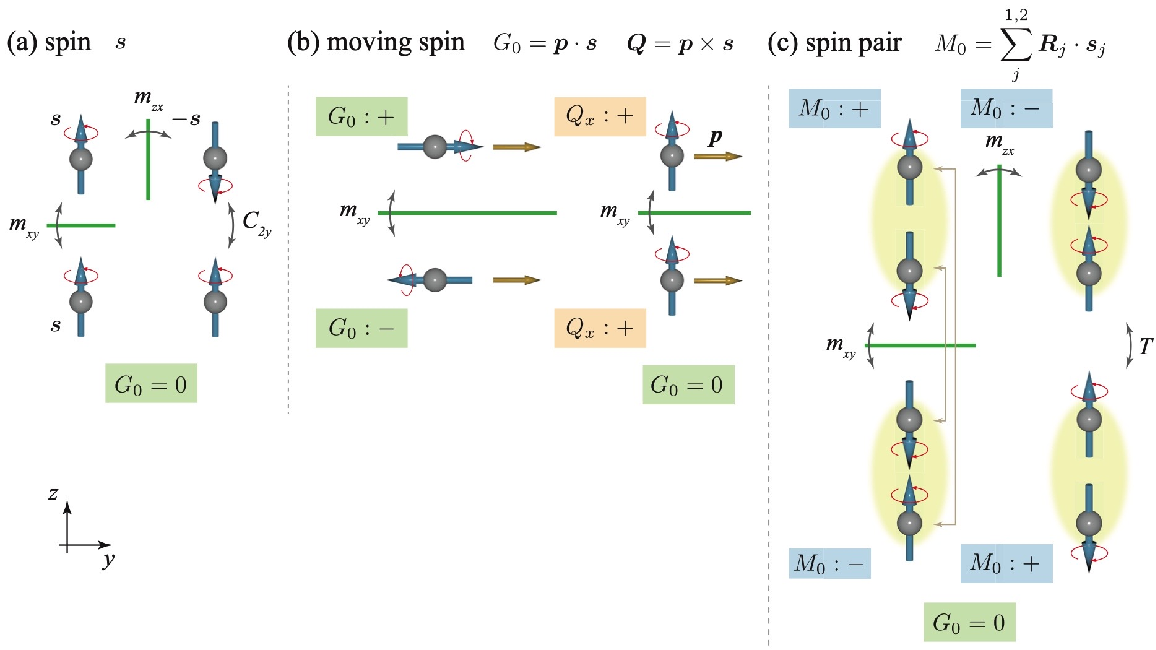}
\vspace{3mm}
\caption{\label{Fig1}
Various spin states under mirror operation.
(a) spin $\bm{s}$ at rest.
Any mirror operations result in the same superimposable spin with an appropriate rotation ($C_{2y}$ denotes $\pi$ rotation along $y$ axis).
(b) spin on moving electron.
Mirror operations give different results depending on the relative directions between spin and momentum $\bm{p}$.
In case of parallel configuration (left), the spin state becomes chiral because its mirror image is not superimposable to one another, while in case of perpendicular configuration (right), mirror image is superimposable, which is characterized by $\bm{Q}$.
(c) anti-parallel spin pair at rest.
Any mirror operations result in the same spin pair by combining it with T operation.
}
\end{figure}

To begin, let us discuss cases involving a single electron.
A single electron at rest (a localized electron) cannot possess $G_{0}$ because any mirror operations ($m_{yz}$, $m_{zx}$, $m_{xy}$) cannot create an enantiomer after combining it with $\pi$ rotation as shown in Fig.~\ref{Fig1}(a).
More precisely, it behaves as a T-odd axial vector.

Meanwhile, a moving electron with a spin $\bm{s}$ parallel or anti-parallel to its momentum $\bm{p}$, as shown in the left panel of Fig.~\ref{Fig1}(b), possesses chirality because it is characterized by helicity $\bm{p}\cdot\bm{s}$ categorized by $G_{0}$.
Indeed, $\bm{p}\cdot\bm{s}$ changes its sign under the P operation, while it remains unchanged under the T operation.
On the other hand, when $\bm{p}$ is perpendicular to $\bm{s}$, as depicted in the right panel of Fig.~\ref{Fig1}(b), this state is achiral ($G_{0}=\bm{p}\cdot\bm{s}=0$) and is characterized by $\bm{Q}=\bm{p}\times\bm{s}$, which behaves as an electric dipole $\bm{Q}$, i.e., a T-even polar vector.
It is notable that any electrons at ground state often have transition matrices to exited states, some of which can possess finite probabilities having chirality.

Next, let us consider two localized electrons with anti-parallel spins.
This pair of spins, viewed as a cluster, break mirror symmetry, as illustrated in Fig.~\ref{Fig1}(c).
In this figure, all $m_{yz}$, $m_{zx}$, $m_{xy}$ operations invert the handedness of the enantiomer.
In this case, outward and inward alignments of spins seem to be named as right-handed and left-handed enantiomers, respectively, for example.
However, what is crucial here is that this spin pair also breaks time-reversal symmetry.
In essence, the T operation can also invert the ``handedness of the enantiomer'', implying a lack of chirality in this context.
In terms of multipole language, such a quantum state is distinctly classified as a magnetic monopole $M_{0}$ (T-odd pseudoscalar), which Barron has referred to as \textit{false} chirality.
While it exhibits a form of chirality according to Kelvin's definition, it does not align with Barron's definition.
Thus, it should be distinguished from \textit{true} chirality denoted as $G_{0}$.

\begin{table}
\caption{\label{Table1}
Properties of monopoles and dipoles, and corresponding examples of materials and fields.
Only the category of T-even pseudoscalar $G_{0}$ matches the definition of chirality.
$M_{\perp}$ and $M_{\parallel}$ denote mirror operations perpendicular and parallel to a dipole vector, respectively.
}
{\centering
\tabcolsep = 2mm
\renewcommand\arraystretch{1.4}
\begin{tabular}{|c|c|c|c|c|c|l|l|} \hline
& Symbol & T & P & $M_{\perp}$ & $M_{\parallel}$ & Example (material) & Example (field) \\ \hline
\multirow{8}{*}{\rotatebox[origin=l]{90}{Monopole}} & $Q_{0}$ & $+$ & $+$ & $+$ & $+$ & Charge & Scalar potential $\phi$ \\ \cline{2-8}
& \multirow{2}{*}{$M_{0}$} & \multirow{2}{*}{$-$} & \multirow{2}{*}{$-$} & \multirow{2}{*}{$-$} & \multirow{2}{*}{$-$} & Anti-parallel spin pair & \multirow{2}{*}{$\bm{E}\cdot\bm{B}$} \\ 
& & & & & & Scalar spin ``chirality'' ($\chi_{ijk}$) & \\ \cline{2-8}
& $T_{0}$ & $-$ & $+$ & $+$ & $+$ & Antiferromagnet in Type I magnetic point group & Kinetic process $t$ \\ \cline{2-8}
& \multirow{5}{*}{\textcolor{ForestGreen}{$G_{0}$}} & \multirow{5}{*}{\textcolor{ForestGreen}{$\bm{+}$}} & \multirow{5}{*}{\textcolor{ForestGreen}{$\bm{-}$}} & \multirow{5}{*}{\textcolor{ForestGreen}{$\bm{-}$}} & \multirow{5}{*}{\textcolor{ForestGreen}{$\bm{-}$}} & \textcolor{ForestGreen}{\bf Chiral molecule/crystal} & \multirow{5}{*}{\textcolor{ForestGreen}{\bf Zilch $\rho_{\chi}$}} \\
& & & & & & \textcolor{ForestGreen}{\bf Helicity ($\bm{p}\cdot\bm{s}$)} & \\
& & & & & & \textcolor{ForestGreen}{\bf Chiral skyrmion (Bloch),\,\,\, Helimagnet} & \\
& & & & & & \textcolor{ForestGreen}{\bf Vector spin chirality ($\bm{\chi}_{ij}$) parallel to the bond} & \\
& & & & & & \textcolor{ForestGreen}{\bf Chiral phonon ($\bm{L}\parallel\bm{k}$)} & \\ \hline
\multirow{7}{*}{\rotatebox[origin=l]{90}{Dipole}} & \multirow{3}{*}{$\bm{Q}$} & \multirow{3}{*}{$+$} & \multirow{3}{*}{$-$} & \multirow{3}{*}{$-$} & \multirow{3}{*}{$+$} & Polar molecule/crystal,\,\,\, Spin current ($\bm{p}\times\bm{s}$) & Electric field $\bm{E}$ \\
& & & & & & Vector spin ``chirality'' ($\bm{\chi}_{ij}$) perpendicular to the bond & Gravity $\bm{F}$ \\
& & & & & & ``Chiral'' phonon ($\bm{L}\perp\bm{k}$) & Displacement $\bm{u}$ \\ \cline{2-8}
& \multirow{2}{*}{$\bm{M}$} & \multirow{2}{*}{$-$} & \multirow{2}{*}{$+$} & \multirow{2}{*}{$+$} & \multirow{2}{*}{$-$} & Ferromagnet,\,\,\,angular momentum ($\bm{L}$, $\bm{s}$) & Magnetic field $\bm{B}$ \\
& & & & & & ``Chiral'' supercondoctor & Rotation $\bm{\omega}$ \\ \cline{2-8}
& \multirow{2}{*}{$\bm{T}$} & \multirow{2}{*}{$-$} & \multirow{2}{*}{$-$} & \multirow{2}{*}{$-$} & \multirow{2}{*}{$+$} & \multirow{2}{*}{Magnetic toroidal (anapole) order} & Electric current $\bm{J}$ \\
& & & & & & & Wave vector $\bm{k}$ \\ \cline{2-8}
& $\bm{G}$ & $+$ & $+$ & $+$ & $-$ & Ferroaxial (rotational) order & Rotational strain $\bm{\nabla}\times\bm{u}$ \\ \hline
\end{tabular}
}
\end{table}

Table~\ref{Table1} shows the properties of four types of monopoles and dipoles for P, T, and mirror operations, and corresponding materials and fields including spin states.
It should be emphasized that the frequently used terms ``chiral'' or ``chirality'' belonging to multipoles other than $G_{0}$ in Table~\ref{Table1} are actually not chiral.
As such an example, the authors will describe so-called scalar/vector spin ``chirality'' which are named in solid state physics without very serious consideration on the definition of chirality.

\begin{figure}
\includegraphics[width=13cm]{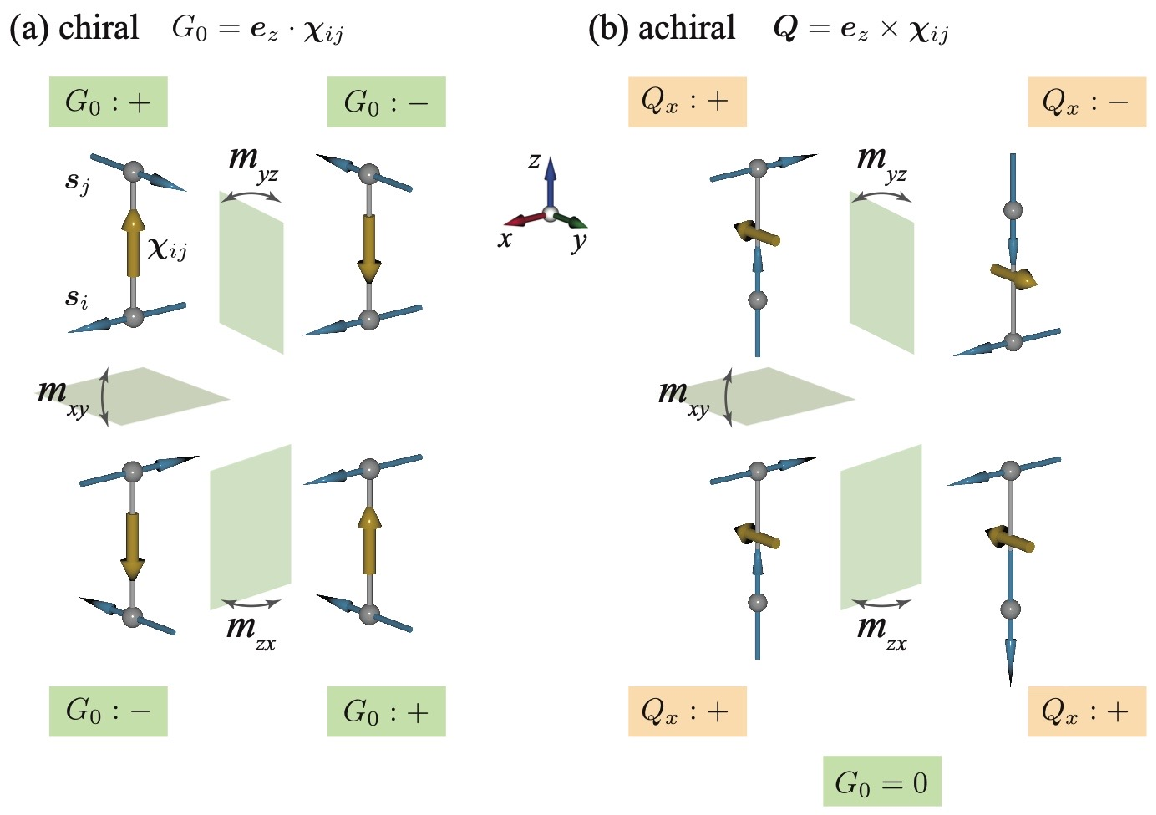}
\caption{\label{Fig2}
Vector spin ``chirality'' (yellow arrows) for (a) parallel to the bond $\bm{e}_{z}$, (b) perpendicular to the bond.
Only the component of $\bm{\chi}$ parallel to the bond is chiral.
}
\end{figure}

The scalar spin \lq\lq chirality'\rq\rq is defined by spins placed on three adjacent lattice points $i$, $j$, $k$, represented as $\bm{\chi}_{ijk}=\bm{s}_{i}\cdot(\bm{s}_{j}\times\bm{s}_{k})$.
This quantity clearly exhibits T-odd and P-odd pseudoscalar characteristics, classified by $M_{0}$.
The condensed state of $\bm{\chi}_{ijk}$ has traditionally been termed as a ``chiral'' spin liquid~\cite{Wen1989}; however, we propose to refer it as a \textit{magnetic monopole liquid} to emphasize its differentiation from the chirality, $G_{0}$.

Next, let us consider the quantity often referred to as vector spin ``chirality'', defined by $\bm{\chi}_{ij}=\bm{s}_{i}\times\bm{s}_{j}$ on an $i$-$j$ bond.
This vector quantity changes its sign under P operation about the bond center and remains invariant under time reversal. Therefore, $\bm{\chi}_{ij}$ is a T-even P-odd vector.
However, each component of this vector behaves differently under mirror operations.
Through explicit mirror operations, as depicted in Fig.~\ref{Fig2}(a), the component of $\bm{\chi}_{ij}$ parallel to the bond direction (taken as the z-axis) changes its sign for all mirror operations, i.e., it is chiral.
On the other hand, for the component of $\bm{\chi}_{ij}$ perpendicular to the bond direction, it behaves as a polar vector, as shown in Fig.~\ref{Fig2}(b).
Hence, we decompose $\bm{\chi}_{ij}$ with respect to the bond direction as follows:
\begin{align}
\bm{\chi}_{ij}=G_{0}\bm{e}_{z}-Q_{y}\bm{e}_{x}+Q_{x}\bm{e}_{y},
\end{align}
where only the component along the bond direction $G_{0}=\bm{e}_{z}\cdot\bm{\chi}_{ij}$ is chiral; otherwise, it behaves
as an electric dipole $\bm{Q}=\bm{e}_{z}\times\bm{\chi}_{ij}$, or achiral.
In summary, equating $\bm{\chi}_{ij}$ with chirality is inappropriate; it should be referred to as chiral only for the component parallel to the bond direction.

\begin{figure}
\includegraphics[width=11cm]{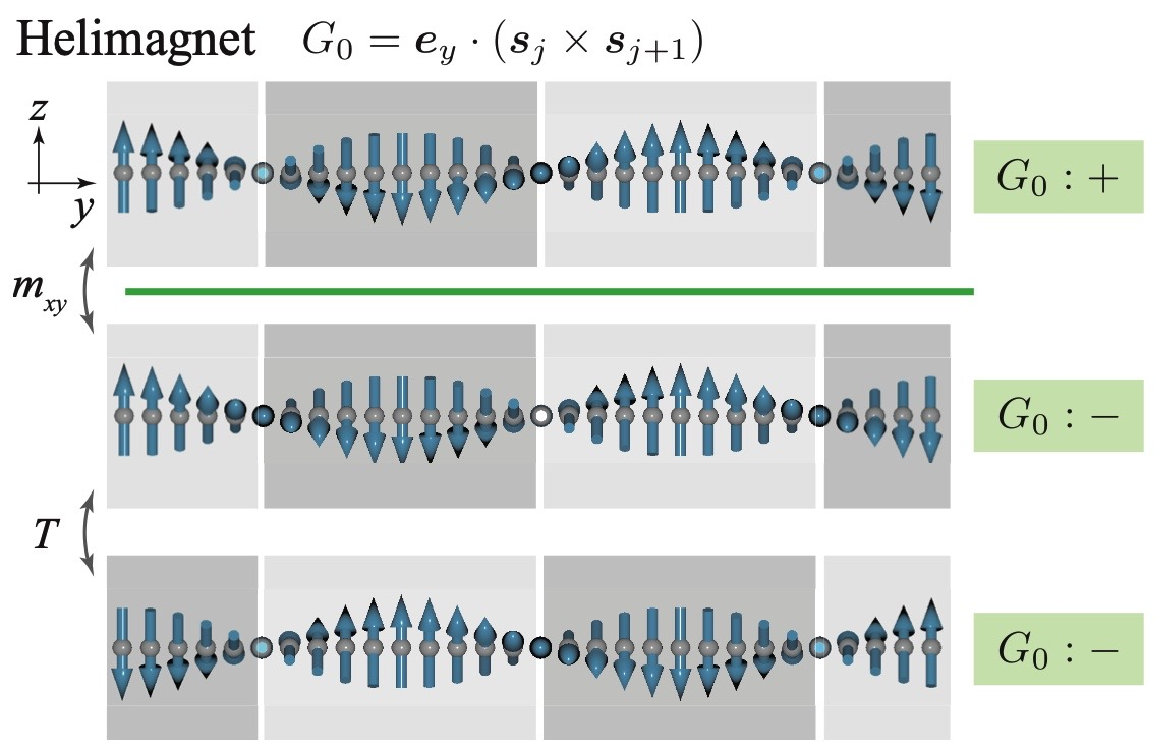}
\caption{\label{Fig3}
Mirror and time-reversal operations for spin configuration of helimagnet.
It behaves as T-even pseudoscalar, i.e., chiral.
}
\end{figure}

By assembling the $G_{0}$ component of vector spin chirality as described above, with lattice translation operations in a crystal lattice, it becomes feasible to produce a helical structure of spins, commonly referred to as helimagnetic order.
This helimagnet also exhibits $G_{0}$ characteristics, as depicted in Fig.~\ref{Fig3}, thus is chiral.
While the T operation flips all the spins, sliding the lattice by a half lattice constant yields precisely the same spin state (denoted by the shaded regions in Fig.~\ref{Fig3}).

\section{Chirality from Dirac equation}

In the preceding section, we categorized various states of materials and fields using multipole language, and elucidated their chirality compatible with its definition, particularly in spin configurations.
To delve deeper into the discussion on the chirality of material, its dynamics, and interactions with surrounding fields, it is imperative to introduce a suitable quantitative measure of chirality for the underlying electronic state.

To address this, let us start with the Dirac equation.
It is well known that for a given 4-component spinor wave function $\psi_{\mu}(x)$, the chirality density, $\chi(x)$, at the space-time point $x=(\bm{r},t)$ is expressed as
\begin{align}
\chi(x)=\sum_{\mu\nu}\psi_{\mu}^{*}(x)\gamma_{\mu\nu}^{5}\psi_{\nu}(x),
\end{align}
where $\gamma^{5}=i\gamma^{0}\gamma^{1}\gamma^{2}\gamma^{3}$ in terms of the Dirac $\gamma$-matrices represents the chirality operator acting on $\psi_{\mu}(x)$.
It is important to note that the helicity, $\bm{p}\cdot\bm{s}$ ($\bm{s}=\hbar\bm{\sigma}/2$), is the independent operator from $\gamma^{5}$.

In the weak relativistic regime, by expanding the inverse of a mass $m$ (or the speed of light $c$), $\psi_{\mu}(x)$ is related to a 2-spinor wave function $\phi_{\alpha}(x)$ ($\alpha=\uparrow,\downarrow$) in the ordinary Schr\"odinger-Pauli equation~\cite{Foldy1950,Gurtler1975,Wang2006,Chen2014}.
The corresponding chirality density in this regime becomes~\cite{Banerjee2020,Abanov2022,Hoshino2023}
\begin{align}
\chi(x)=\sum_{\alpha\beta}\phi_{\alpha}^{*}(x)\frac{\bm{\pi}\cdot\bm{\sigma}_{\alpha\beta}}{mc}\phi_{\beta}(x),
\quad
(\bm{\pi}=\bm{p}+\frac{e}{c}\bm{A}),
\end{align}
where $\bm{p}$, $\bm{\sigma}$, $-e$, and $\bm{A}$ denote the canonical momentum operator, Pauli matrices, electron charge, and vector potential, respectively.
In the following, we consider the case of no external electromagnetic field, $\bm{A}=0$.
It is worth noting that the operator $\bm{\pi}\cdot\bm{\sigma}$ has a T-even pseudoscalar property, and there is no essential distinction between chirality and helicity in the weak relativistic regime.

When a system is chiral, the wave function $\phi_{\alpha}(x)$ is characterized by having finite expectation value of time-averaged $\chi(x)$, i.e., $\chi =\frac{1}{T}\int_{0}^{T}dt\int d\bm{r}\chi(\bm{r},t)$, where $T$ is sufficiently
long time interval.
The sign of $\chi$ represents its handedness, in other words, right-handed (R) and left-handed (L) systems are related with each other by $\chi_{\rm R}=-\chi_{\rm L}$.
In the case of structural chirality with two enantiomers, $\chi_{\rm R}$ and $\chi_{\rm L}$ correspond to each handedness~\cite{Oiwa2022,Inda2024}.
It should be emphasized that the present measure of chirality is capable to distinguish a pair for non-structural chirality as well.

The chirality density operator can be expressed in the multipole language in the spherical coordinate.
By using the mathematical identity for $\bm{p}$ operator in terms of the orbital angular-momentum operator $\bm{l} = \bm{r}\times\bm{p}$ as
\begin{align}
\bm{p}=-\frac{1}{r^{2}}\left[(\bm{r}\times\bm{l})+i\hbar\bm{r}r\frac{\partial}{\partial r}\right],
\end{align}
we obtain the chirality density operator in the multipole language as
\begin{align}
\frac{\bm{p}\cdot\bm{\sigma}}{mc}=-\frac{1}{mcr^{2}}\left[\bm{r}\cdot(\bm{l}\times\bm{\sigma})+i\hbar(\bm{r}\cdot\bm{\sigma})r\frac{\partial}{\partial r}\right].
\end{align}
This equation tells us that, since chirality is a non-local concept, its definition requires some \textit{extent}; namely, it is expressed by expanding the relevant \textit{space} to include spin space (first term) or by incorporating derivatives of real space (second term).
In this sense, chirality has internal and/or geometrical structures.

Here, the chirality density operator consists of two types ET monopoles, $G_{0}^{(1)}$ and $G_{0}^{(2)}$, which are proportional to $\bm{r}\cdot(\bm{l}\times\bm{\sigma})$ and $i(\bm{r}\cdot\bm{\sigma})$.
From these expressions, the relation to other multipoles is apparent: $\bm{G}=(\bm{l}\times\bm{\sigma})$ and $M_{0}=(\bm{r}\cdot\bm{\sigma})$ are the ET dipole and the M monopole.
$\bm{G}$ and $M_{0}$ have T-even axial vector and T-odd pseudoscalar properties, respectively, as shown in Table~\ref{Table1}, and they are totally independent operators from $G_{0}^{(1)}$ and $G_{0}^{(2)}$.
It is interesting to note that they can become a source of $G_{0}$ after combining with other type of multipoles.
In particular, $M_{0}$ being odd for all the mirror operations differs from $G_{0}$ in time-reversal property as discussed in the previous section.
Nevertheless, $M_{0}$ can still be converted to $G_{0}$ by time-domain transformation.
In other words, pure imaginary number corresponds to T-odd scalar (i.e. MT monopole, $T_{0}$) as the time derivative or integration gives $-i\omega$ or $-1/i\omega$ in the frequency domain.
Moreover, from the symmetry point of view, $\bm{p}$ or wave vector $\bm{k}$ has common property of $i\bm{r}$.
Similarly, the magnetic field $\bm{B}$ has common property of $\bm{\sigma}$.
Therefore, there are symmetry correspondences,
\begin{align}
G_{0}\leftrightarrow\bm{r}\cdot\bm{G}\leftrightarrow iM_{0}\leftrightarrow \bm{k}\cdot\bm{B},
\quad\text{etc.},
\end{align}
These relations are essential to consider the emergence of chirality by multipole interconversion.

Let us illustrate a specific example of chiral (ET-monopole) states of an electron by considering four states composed of $s$ and $p$ wave functions with spins within the total angular momentum $j = 1/2$ manifold.
One eigenstate of $G_{0}^{(1)}$ with a positive eigenvalue in the spherical coordinate is given by
\begin{align}
\phi_{+;a}(\theta,\phi)=\frac{1}{\sqrt{2}}
\begin{pmatrix}
1-i\cos\theta \\ -ie^{i\phi}\sin\theta
\end{pmatrix},
\end{align}
in two-component spinor form, and its T partner,
\begin{align}
\phi_{+;b}(\theta,\phi)=\frac{1}{\sqrt{2}}
\begin{pmatrix}
-ie^{-i\phi}\sin\theta \\ 1+i\cos\theta
\end{pmatrix},
\end{align}
has the same eigenvalue because a chiral state does not break T symmetry, which is obtained by applying $-i\sigma_{y}$ and complex conjugation for $\phi_{+;a}(\theta,\phi)$.

On the contrary, the P partners of $\phi_{+;a}$ and $\phi_{+;b}$ have eigenvalues with opposite sign.
They are given by
\begin{align}
\phi_{-;a}(\theta,\phi)=\frac{1}{\sqrt{2}}
\begin{pmatrix}
1+i\cos\theta \\ ie^{i\phi}\sin\theta
\end{pmatrix},
\quad\text{and}\quad
\phi_{-;b}(\theta,\phi)=\frac{1}{\sqrt{2}}
\begin{pmatrix}
ie^{-i\phi}\sin\theta \\ 1-i\cos\theta
\end{pmatrix}.
\end{align}
Note that P operation is given by $\theta\to\pi-\theta$ and $\phi\to\phi+\pi$.
Therefore, $\phi_{+;a/b}$ and $\phi_{-;a/b}$ constitute a pair of enantiomers.
This is a prime example of non-structural chirality.

\section{Emergence of chirality by multipole interconversion}

Chirality characterized by $G_{0}$ can arise not just from electronic states but also from physical fields or their combinations.
Optical chirality, as exemplified by Lipkin's Zilch~\cite{Lipkin1964,Proskurin2017}, is one such case, defined by
\begin{align}
\rho_{\chi}=\bm{E}\cdot(\bm{\nabla}\times\bm{E})+\bm{B}\cdot(\bm{\nabla}\times\bm{B}).
\end{align}
Depending on the handedness of circularly polarized light (CPL), $\rho_{\chi}$ can be positive or negative, with degenerate electromagnetic field energies.

Another example is the scalar product between electrical current $\bm{J}$ and magnetic field $\bm{B}$, which is also categorized by $G_{0}=\bm{J}\cdot\bm{B}$, because this quantity is inverted by P ($\bm{J}\to-\bm{J}$ and $\bm{B}\to\bm{B}$), but not by T ($\bm{J}\to-\bm{J}$ and $\bm{B}\to-\bm{B}$).
Both positive and negative $G_{0}$ have degenerate energy, but they can be interchanged simply by inverting either $\bm{J}$ or $\bm{B}$.
It is worth noting that while the electrical current $\bm{J}$ is categorized in the MT dipole $\bm{T}$, it can transform 
into $G_{0}$ under the influence of magnetic field $\bm{B}$ categorized by the magnetic dipole $\bm{M}$.
Note that electric field $\bm{E}$ is categorized by $\bm{Q}$ and thus it can transform to $M_{0}$ when combined with $\bm{B}$.

It is also intriguing to consider the emergence of chirality from ferro-axial materials characterized by the ET dipole $\bm{G}$, when subjected to a static electric field along $\bm{G}$~\cite{Hayashida2021,Fang2023}.
In such cases, the handedness of the resulting chiral material can be switched between right and left by selecting the direction of $\bm{E}$ in a single-domain ferro-axial state.
Both right and left states possess degenerate energy.
This process represents an interconversion from $\bm{G}$ to $G_{0}=\bm{G}\cdot\bm{E}$ facilitated by $\bm{E}$.
Such chirality switching holds potential future applications in information processing based on chirality.
Typically, structural chirality switching using external fields over a short period is challenging.
However, this method of $G_{0}$ generation enables high-speed alternation of chirality, which could be utilized, for instance, in spin polarization switching through CISS effects.

Another significant interconversion regarding chirality involves transitions between $G_{0}$ and $M_{0}$.
Despite the presence of several experimental scenarios demonstrating this interconversion, its microscopic mechanism remains debated.
For instance, in 2012, researchers demonstrated that $M_{0}$ generated by effective gravity and rotation can lead to the formation of a chiral molecular aggregate $G_{0}$ through a kinetic process~\cite{Micali2012}.

\begin{figure}
\includegraphics[width=14cm]{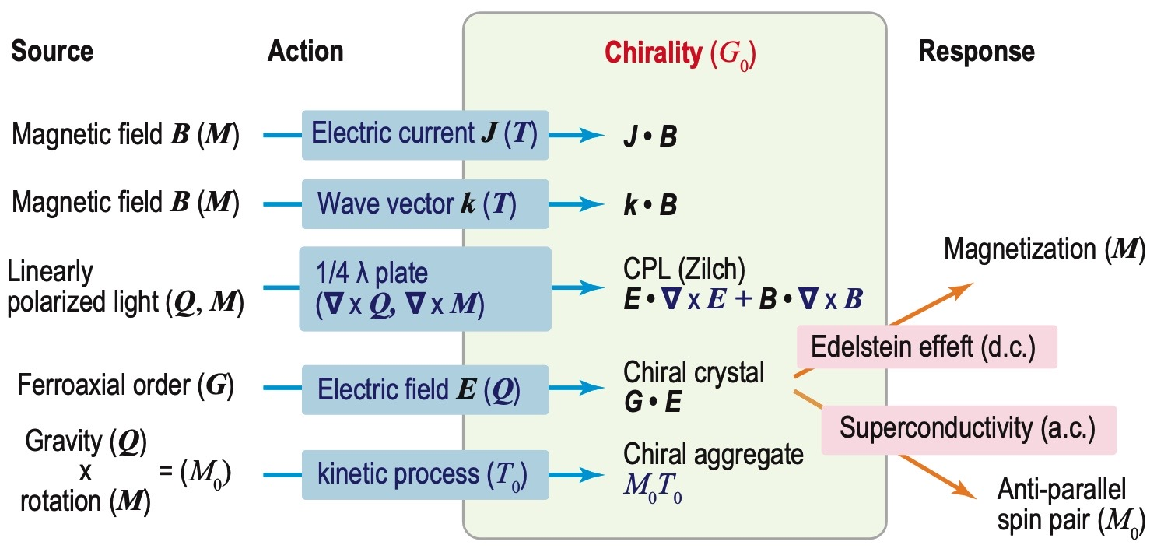}
\caption{\label{Fig4}
Emergence of chirality from other multipoles, and its interconversion to different multipoles.
Symbols in the parentheses denote categories of multipoles.
}
\end{figure}

In equilibrium or in the ground state, such an $M_{0}$ exists as a completely independent state from chirality $G_{0}$.
However, under non-equilibrium conditions, it appears that $G_{0}$ can serve as a source of $M_{0}$ through integration or differentiation in the time domain, corresponding to $T_{0}$.
One of the authors has also experimentally validated that an anti-parallel spin pair ($M_{0}$) can be produced by exciting the superfluid with structural chirality in a superconductor~\cite{Nakajima2023}.
Of particular interest is the one-to-one correspondence between the signs of $M_{0}$ and $G_{0}$ in the interconversion process in this experiment.
As a result, it becomes possible to determine the handedness of the chiral material solely by observing the inward or outward orientation of the anti-parallel spin pair.
Various other type of interconversions between different multipoles and the resultant form of $G_{0}$ (chirality) are outlined in Fig.~\ref{Fig4}.

\section{Enantiospecific responses in chiral materials}

\begin{figure}
\includegraphics[width=12cm]{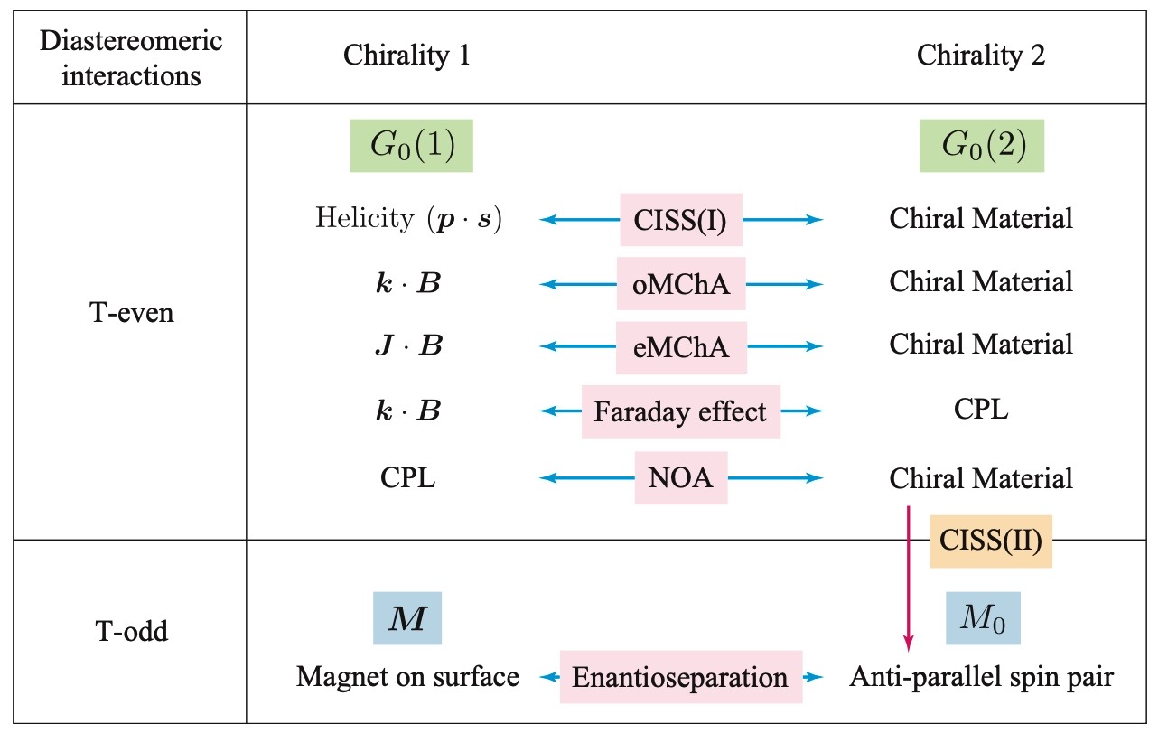}
\caption{\label{Fig5}
Diastereomeric interactions among different chiral entities.
}
\end{figure}

Once $G_{0}$ is generated by materials and/or physical fields, we can consider scenarios where two or more chiral states coexist and interact with each other.
One such example is termed electrical magneto-chiral anisotropy (eMChA)~\cite{Rikken2001,Pop2014}, which has become a current hot topic in solid-state physics.
In the case of eMChA, the combination of electrical current $\bm{J}$ and collinear magnetic field $\bm{B}$ can form $G_{0}(1)=\bm{J}\cdot\bm{B}$, which interacts with material chirality $G_{0}(2)$ ($G_{0}(1)$-$G_{0}(2)$ interaction; see also the reference [\onlinecite{Kishine2022}] and Fig.~\ref{Fig5}).
Through this interference, the electrical resistances for positive and negative $G_{0}(1)$ states become different for a given enantiopure chiral conductor characterized by $G_{0}(2)$, which is nothing but the nonreciprocal transport.
In such cases, the energies of the two $G_{0}(2)$ states (right- or left-handedness) are no longer degenerate.

This interpretation is clearly illustrated for optical MChA (oMChA)~\cite{Atzori2021}, by expanding the dielectric tensor with respect to the wave vector $\bm{k}$ of light and $\bm{B}$ as
\begin{align}
\varepsilon_{ij}=Q_{0}\delta_{ij}+Q_{ij}+G_{ijkl}k_{k}B_{l},
\end{align}
where the components $Q_{0}$, $Q_{ij}$, and $G_{ijkl}$ are categorized by E monopole, E quadrupole, and ET hexadecapole of material, respectively.
In the simplest case of a cubic chiral system, $Q_{ij}=0$ and $G_{ijkl}=G_{0}(2)\delta_{ij}\delta_{kl}$, and thus the dielectric tensor becomes diagonal, $\varepsilon=Q_{0}+G_{0}(2)(\bm{k}\cdot\bm{B})$.
Indeed, $G_{0}(2)$ of material interacts with another chirality, $G_{0}(1)=\bm{k}\cdot\bm{B}$, which gives an anisotropic response depending on $\bm{k}$ and $\bm{B}$.

\begin{figure}[b]
\includegraphics[width=8cm]{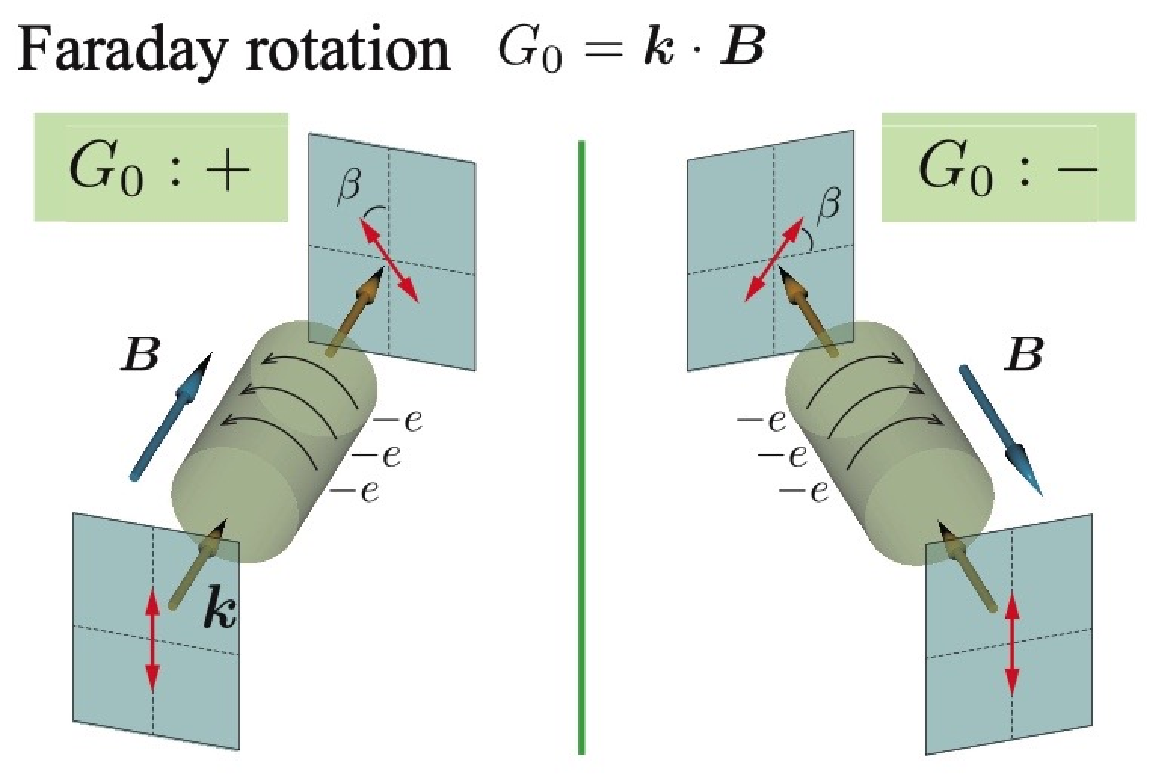}
\caption{\label{Fig6}
Right-handed and left-handed Faraday rotation.
}
\end{figure}

It is also noteworthy that the Faraday configuration of polarized light rotation in the magnetic field $\bm{B}$ or the internal molecular field in the case of ferromagnets is understood through the concept of $G_{0}$.
The optical rotation arises from the anti-symmetric off-diagonal component of the dielectric tensor, $\varepsilon_{ij}=\epsilon_{ijk}G_{k}$, where $\bm{G}$ represents a pure imaginary ET dipole and is given by $iQ_{0}\bm{B}$ in isotropic material.
By introducing the projection of $\bm{G}$ onto the direction of light propagation ($\bm{k}/|\bm{k}|$, considered as the z-axis) as $g=-i\bm{k}\cdot\bm{G}/|\bm{k}|$, the dielectric tensor (with only $xy$ components shown) becomes:
\begin{align}
\tilde{\boldmath{\varepsilon}}=
\begin{pmatrix}
\varepsilon_{n} & ig \\ -ig & \varepsilon_{n}
\end{pmatrix}.
\end{align}
For this dielectric tensor, the eigenmode of light propagation in the material is right-handed ($+$) or left-handed ($-$) circularly polarized light (CPL), and its dielectric constant is given by
\begin{align}
\varepsilon_{\pm}=\varepsilon_{n}\pm Q_{0}g
=\varepsilon_{n}\pm Q_{0}G_{0}(1)/|\bm{k}|,
\quad
G_{0}(1)=\bm{k}\cdot\bm{B},
\end{align}
which indicates that CPL ($\pm$ sign) regarded as another chirality $G_{0}(2)$ interacts with the chirality $G_{0}(1)$ differently depending on the sign of $G_{0}(1)$ in the Faraday setup.
Note that linearly polarized light is nothing but a racemic mixture of ($+$)-CPL and ($-$)-CPL.
As a result, the optical rotation angle is proportional to the difference of $\varepsilon_{+}$ and $\varepsilon_{-}$, i.e., $g\propto G_{0}(1)$, and it is inverted when $\bm{B}$ is opposite, as shown in Fig.~\ref{Fig6}.
In this sense, it is legitimate to say the Faraday rotation is a chiral phenomenon, but the effect does not require any chiral material.
The authors would like to point out that there is confusion in understanding the chirality of the Faraday effect because historically, the Faraday effect is often regarded simply as achiral.
However, the correct way to describe it should be that \textit{the Faraday setup is chiral in the sense that the mirror image of the setup, characterized by $G_{0}(1)=\bm{k}\cdot\bm{B}$, is not superimposable to one another, but the effect does not require any chiral material to couple with}.

The natural optical activity (NOA) is understood in a similar manner as follows.
In a gyrotoropic material, the ET vector $\bm{G}$ is given by $G_{i}=iG_{ij}'k_{j}$, and the gyrotropic tensor $G_{ij}'$ can be decomposed as
\begin{align}
G_{ij}'=G_{0}(2)\delta_{ij}+\epsilon_{ijk}Q_{k}+G_{ij},
\end{align}
where the coefficients $G_{0}(2)$, $Q_{k}$, $G_{ij}$ are categorized by ET monopole, E dipole, and ET quadrupole of material, respectively.
In this case, the chiral quantity appears in $g=(G_{0}(2)\bm{k}^{2}+G_{ij}k_{i}k_{j})/|\bm{k}|$.
When only $Q_{k}$ exists, it is called \textit{weak gyrotoropic}, otherwise \textit{strong gyrotoropic}, since the former gives no optical rotation.
It is also important to note that achiral but gyrotoropic material, i.e., $G_{0}(2)=0$ and $G_{ij}\ne0$, also exhibits NOA.

\begin{figure}
\includegraphics[width=10cm]{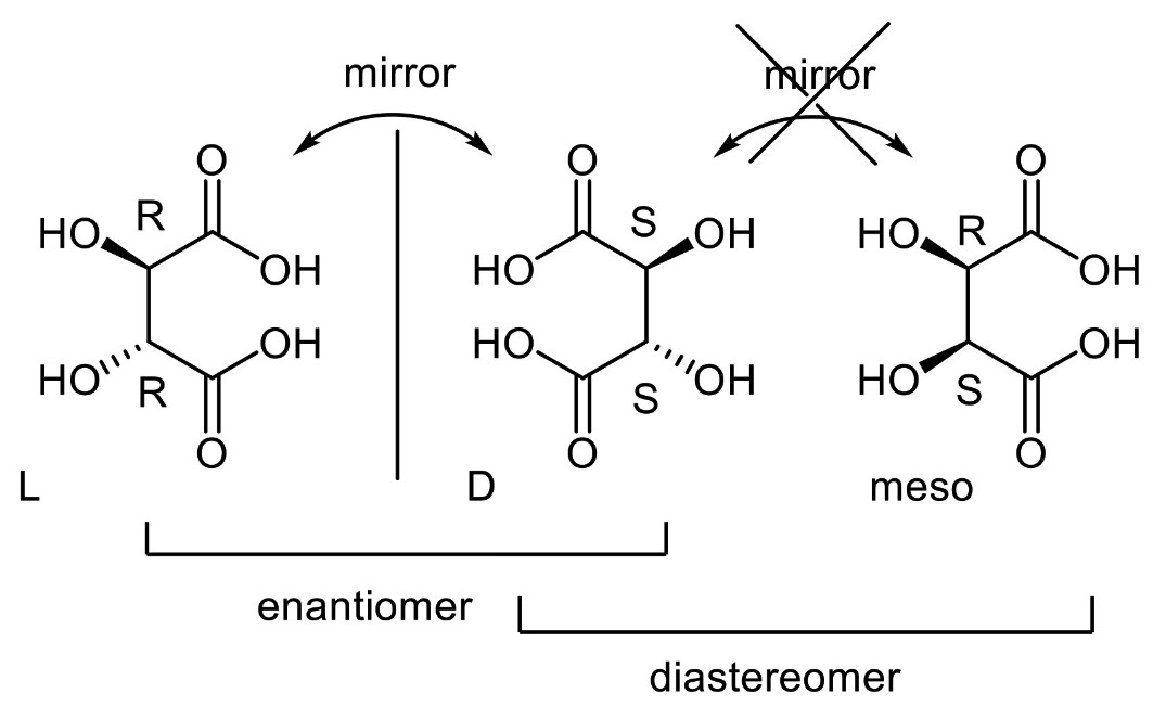}
\caption{\label{Fig7}
Enantiomers (L and D tartaric acid) and diastereomers (D and meso-tartaric acid).
The two asymmetric carbons in L tartaric acid have (R, R) configuration which is a completely mirror-reflected image of (S, S) configuration in D tartaric acid.
On the other hand, two asymmetric carbons in meso form have (R, S) configuration, which will give different energy and physical properties than D/L form because they are not related by mirror operation.
}
\end{figure}

In the field of chemistry, a pair of isomeric molecules that comprise two or more chiral elements (such as asymmetric carbons) but do not form a mutual enantiomeric relationship are called diastereomers.
Diastereomeric molecules differ in various physical properties such as energy, dipole moment, and permittivity because they are not mirror images of one another, as illustrated in Fig.~\ref{Fig7}, for example.

\begin{figure}[t]
\includegraphics[width=12cm]{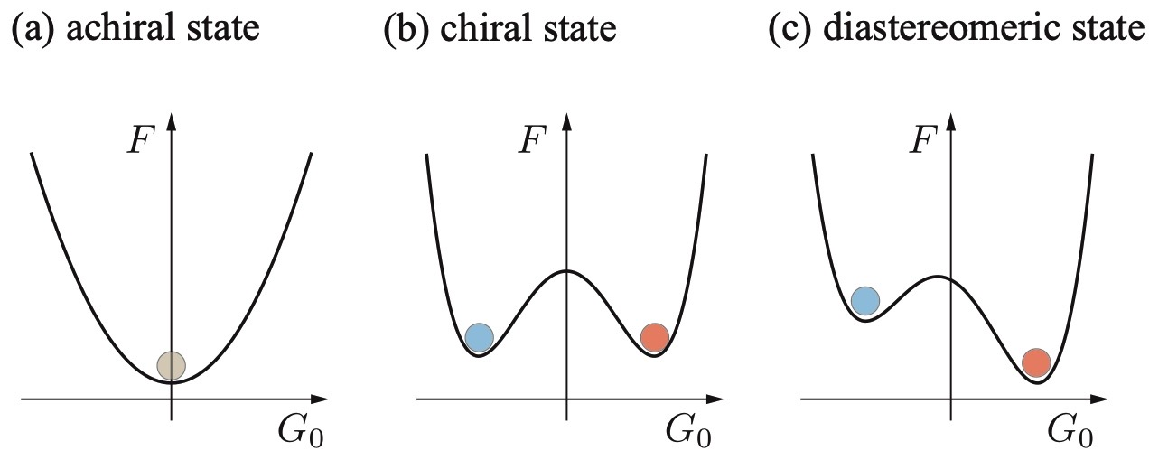}
\caption{\label{Fig8}
Landau free energy in (a) achiral (disordered) state, (b) chiral (ordered) state, and (c) diastereomeric state, which corresponds to the case that the order parameter $G_{0}(1)$ couples with its conjugate (molecular chiral) field $G_{0}(2)$.
}
\end{figure}

From a physics perspective, this situation is understood in the context of the free energy landscape in the Landau theory, as depicted in Fig.~\ref{Fig8}.
As an analogy to the diastereomeric interaction between two chiralities, we can now discuss the relationship between diastereomeric state and three types of states in the Landau theory (Fig.~\ref{Fig8}).
If $G_{0}(1)$ of a target system interacts with other $G_{0}(2)$, the energy degeneracy of two chiral states (positive or negative $G_{0}(1)$) in the target is lifted, and the chiral states should respond differently depending on its handedness.
This is what is called circular dichroism in a broader sense.
For example, CPL can distinguish chirality of materials because the optical chirality (Lipkin's Zilch) possesses $G_{0}(2)$ that can interact with material chirality $G_{0}(1)$.
Therefore, the absorbance coefficients are different between right-handed and left-handed materials.

CISS is another interesting example of diastereomeric interactions between two chiral objects. In this case, special attention is needed because there are two different cases for CISS, which can be tentatively called CISS(I) and CISS(II).

A typical example of CISS(I) is the spin polarization detected in photoemission~\cite{Ghler2011} or bulk chiral crystal experiments~\cite{Inui2020,Shishido2021,Ohe2024}.
In this case, the electron helicity $G_{0}(1)$ interacts with the material chirality $G_{0}(2)$ in a diastereomeric manner to yield the spin polarization parallel to or anti-parallel to the momentum.
CISS(II) is more complicated because it necessitates assuming an anti-parallel spin pair on the molecular edges~\cite{BanerjeeGhosh2018,Waldeck2021}.
In this particular case, the molecular chirality $G_{0}$ is converted into $M_{0}$ as mentioned in the above discussion.
What is important here is the correspondence between the signs of $G_{0}$ and $M_{0}$ during the conversion.
Experiments with chiral superconductors and enantioseparation on magnetic substrates have shown that the signs of $G_{0}$ and $M_{0}$ are correlated with each other.
This correlation indicates that the inward/outward anti-parallel spin pair on the material edges ($M_{0}$) reflected the molecular chirality ($G_{0}$).
The transformation of chiral information into a T-broken state is a novel concept in materials science and appears to be expandable in more diverse ways.

\section{Summary}

In this paper, the authors explored various aspects of chirality, encompassing both theoretical frameworks and experimental observations across physics and chemistry.
We began our discussion by examining the behavior of a single electron spin at rest and in motion, elucidating the meaning of chirality with respect to mirror, inversion, and time-reversal operations.
The distinction between true and false chirality is highlighted, emphasizing the significance of $G_{0}$ (T-even pseudoscalar) in spin configurations.
We also highlighted the emergence of chirality by physical fields via multipole interconversions.
Chirality arising from combination of material quantities and/or physical fields, such as optical chirality $\rho_{\chi}$ and $\bm{J}\cdot\bm{B}$, is discussed.
The paper delves into the interconversion between different types of multipoles and $G_{0}$, shedding light on the mechanisms behind the emergence of chirality in various materials.
The interaction among chiralities made from materials and physical fields was also examined.
Examples such as CISS, MChA, and NOA illustrate how chirality influences the responses of chiral materials, leading to distinct physical properties and behaviors.

Throughout this paper, an intimate relationship between chirality and other multipoles, especially spin configuration, has been found.
Therefore, chirality, which could be a binary information carrier, is expected to work as a source of spin polarizer for the purpose of spintronics and spin-related chemistry.
The authors believe that these fundamental aspects of chirality are quite important in utilizing it in applied physics and chemistry.
One of the authors has recently developed a switchable magnetoresistance device with CISS effect using a chiral molecular motor.
In this device, handedness of the molecular structure could be controlled by external stimuli such as light and heat~\cite{Suda2019}.
However, it took around five minutes to switch the molecular chirality in that case.
In order to find faster chirality switching in molecules, one may need smaller energy barrier between enantiomers, but it will racemize the molecule under thermal fluctuation in general, and the information will be lost.
On the other hand, emergent chirality generated by a combination of several spins, material states, and physical fields could provide more stable, more efficient, and more controllable chiral information systems.
In the quest for chirality-based spin science, the researchers may also encounter dynamics or fluctuation of chirality where time dependence of $G_{0}$ becomes significant.
Such fluctuation induces a coupling between electron and phonon $G_{0}$s for example, which may enhance together some kinds of emergent effects, which will be a valuable opportunity to further develop chirality-based technology.
We hope that the present paper provides an overview of chirality across different research fields, highlighting its fundamental importance and potential applications in diverse fields of science and technology.

\begin{acknowledgments}
The authors would like to thank the members of Quantum Research Center for Chirality (QuaRC), Y. Togawa, T. Satoh, and Y. Kato for stimulating discussions.
We are also grateful to S. Hayami and R. Oiwa for fruitful discussions on multipole interconvesions.
This work was supported by JSPS KAKENHI Grants Nos.~23H00291, 21H01032, 21H01034, 23K03288, 23K20825, 23H00091, and Cooperative Research by Institute for Molecular Science (IMS program 22IMS1220).
This research was also supported by the grant of OML Project by the National Institutes of Natural Sciences (NINS program No.~OML012301).
\end{acknowledgments}

\nocite{*}
\bibliographystyle{aipnum4-2}
\bibliography{ref}

\end{document}